# Fingerprints of the Strong Interaction between Monolayer MoS$_2$ and Gold

Matěj Velický,*[1,2,3] Alvaro Rodriguez,[4] Milan Bouša,[4] Andrey V. Krayev,[5] Martin Vondráček,[6] Jan Honolka,[6] Mahdi Ahmadi,[2] Gavin E. Donnelly,[3] Fumin Huang,[3] Héctor D. Abruña,[2] Kostya S. Novoselov,[1,7,8] and Otakar Frank*[4]

[1] Department of Physics and Astronomy, University of Manchester, Oxford Road, Manchester, M13 9PL, United Kingdom

[2] Department of Chemistry and Chemical Biology, Cornell University, Ithaca, New York, 14853, United States

[3] School of Mathematics and Physics, Queen's University Belfast, University Road, Belfast, BT7 1NN, UK

[4] J. Heyrovský Institute of Physical Chemistry, Czech Academy of Sciences, Dolejškova 2155/3, 182 23 Prague, Czech Republic

[5] HORIBA Scientific, Novato, CA, 94949, United States

[6] Institute of Physics, Czech Academy of Sciences, Na Slovance 1999/2, 182 21 Prague 8, Czech Republic

[7] Centre for Advanced 2D Materials, National University of Singapore, 117546, Singapore

[8] Chongqing 2D Materials Institute, Liangjiang New Area, Chongqing, 400714, China

**ABSTRACT**

Gold-mediated exfoliation of $MoS_2$ has attracted considerable interest in the recent years. A strong interaction between $MoS_2$ and Au facilitates preferential production of centimeter-sized monolayer $MoS_2$ with near-unity yield and provides a heterostructure system noteworthy from a fundamental standpoint. However, little is known about the detailed nature of the $MoS_2$–Au interaction and its evolution with the $MoS_2$ thickness. Here, we identify specific vibrational and binding energy fingerprints of such strong interaction using Raman and X-ray photoelectron spectroscopy, which indicate substantial strain and charge-transfer in monolayer $MoS_2$. Near-field tip-enhanced Raman spectroscopy reveals heterogeneity of the $MoS_2$–Au interaction at the nanoscale, reflecting the spatial non-conformity between the two materials. Far-field micro-Raman spectroscopy shows that this interaction is strongly affected by the roughness and cleanliness of the underlying Au. Our results elucidate the nature of the strong $MoS_2$–Au interaction and provide guidance for strain and charge doping engineering of $MoS_2$.

Several groups have recently introduced a method of exfoliating large-area transition metal dichalcogenides (TMDCs) monolayers using gold substrates or sacrificial layers.[1-3] For $MoS_2$ in particular, it is possible to prepare centimeter-sized monolayers due to their preferential, near-unity exfoliation yield and the high quality of natural molybdenite.[3] This is a major advancement for mechanical exfoliation, which produces the highest-quality crystals but has been challenging to scale up, unlike the readily scalable chemical vapor deposition or liquid phase-exfoliation that produce lower quality crystals. Gold-mediated mechanical exfoliation has quickly attracted attention and has been utilized in fabrication of flexible gas sensors,[4] lithography patterning for transistor applications,[5] and construction of large-area vdW heterostructures.[6,7] A significant advantage of this method is the polymer-free post-transfer of the gold-exfoliated TMDCs, which leaves their surfaces free from residual contamination.[2,8]

When bulk $MoS_2$ is pressed against freshly-deposited Au and peeled off, monolayer $MoS_2$ crystals with a near-unity yield remain on the Au surface.[2,3] This was rationalized theoretically, showing that the binding energy between the bottom-most $MoS_2$ layer and Au is larger than the interlayer equivalent in bulk $MoS_2$,[3] and that both tensile and compressive biaxial strains induced in $MoS_2$ facilitate preferential monolayer exfoliation in the naturally AB-stacked molybdenite (2H phase in Ramsdell notation).[9] Cross-sectional scanning transmission electron microscopy and X-ray photoelectron spectroscopy (XPS) confirmed that the $MoS_2$–Au interaction is of van der Waals (vdW) rather than covalent nature.[3] This was corroborated by spectroscopic and electronic studies of gold-exfoliated $MoS_2$ with the Au removed, which showed that the Raman, photoluminescence, and field-effect transistor responses were qualitatively identical to those of the semiconducting 1H phase of monolayer $MoS_2$ exfoliated directly onto insulating substrates.[1,2,5,6] These observations signify that the $MoS_2$ metallicity endowed by the Au[3,10] can be reversed after transfer onto another substrate, significantly increasing the scope of this method to optoelectronics, photovoltaics, and photocatalysis.

Despite these research efforts, little is known about the nature of the strong interaction between $MoS_2$ and Au, its dependence on the number of $MoS_2$ layers, and specific effects that the Au brings about in $MoS_2$. This is most likely due to the lack of reliable measurements directly on the Au substrate. In this study, we reveal the spectroscopic fingerprints of the strong monolayer $MoS_2$–Au interaction, using Raman spectroscopy and XPS of $MoS_2$ exfoliated on a range of Au substrates, prepared by magnetron sputtering, electron-beam (e-beam) evaporation, and thermal evaporation. Far-field micro-Raman and micro-XPS of mono- and few-layer $MoS_2$ on Au reveal additional peaks with differing vibrational frequencies and binding energies, respectively, compared to $MoS_2$ on an insulating substrate, which are explained by an Au-induced strain and charge doping of the bottom-most $MoS_2$ layer. These results also suggest heterogeneity of the $MoS_2$–Au interaction, which is unequivocally confirmed by near-field tip-enhanced Raman spectroscopy (TERS) with 10 nm spatial resolution. Finally, we observe clear correlations of the $MoS_2$–Au interaction strength with the roughness and cleanliness of the underlying Au, which originate in non-conformality between the two materials.

*Micro-Raman Spectroscopy and Micro-XPS of MoS₂ on Au*

Monolayer (1L), bilayer (2L), trilayer (3L), and bulk MoS$_2$ crystals were readily identified due to their high optical contrast (Fig. 1a).[11] The far-field micro-Raman spectra of 1L MoS$_2$ on Au exhibit conspicuous broadening and downshift of the E′ mode and splitting of the A$_1$′ mode, in comparison to 1L MoS$_2$ on SiO$_2$/Si (Fig. 1b). Lattice deformation (strain) and charge-transfer (doping) are the two main factors influencing Raman frequencies in 1L MoS$_2$.[12] The effect of strain is more pronounced for the in-plane E′ phonon,[13] while carrier doping has a greater influence on the out-of-plane A$_1$′ phonon.[14,15] The E′ mode broadening and downshift can thus be interpreted as heterogeneous biaxial strain, originating in the lattice mismatch between MoS$_2$ and Au.[16] The induced change in the frequency of a generic Raman mode M can be estimated as $\delta\omega_M = \omega_M^0 - \omega_M = 2\gamma_M \omega_M^0 \varepsilon$, where $\omega_M^0$ and $\omega_M$ are the Raman frequencies of the M mode in unstrained and strained lattices, respectively, $\gamma_M$ is the Grüneisen parameter of the M mode, and $\varepsilon$ is the biaxial strain.[12]

Since the precise values of zero-strain Raman frequencies in 1L MoS$_2$ ($\omega_{E'}^0$ and $\omega_{A_1'}^0$) are unknown, we use $\omega_{E'}^{SiO_2/Si} = (385.9 \pm 0.2)$ cm$^{-1}$ measured on SiO$_2$/Si as a reference. The E′ mode peak frequency for all the 1L MoS$_2$/Au samples in this study averages at $\omega_{E'} = (378.2 \pm 0.6)$ cm$^{-1}$, which yields $\delta\omega_{E'} = 7.7$ cm$^{-1}$ and implies a tensile strain of $\varepsilon = (1.2 \pm 0.1)\%$ when $\gamma_{E'} = 0.82$ is used (average from refs [12,13]). If the observed broadening of the E′ mode on Au with a linewidth of $\Gamma_{E'} = (6.1 \pm 0.5)$ cm$^{-1}$ were caused solely by heterogeneous lattice deformation, the biaxial tension would fall between 0.6% and 1.9%, taken as 5$^{th}$ and 95$^{th}$ percentiles of the distribution of peaks with $\Gamma_{E'} = 2.4$ cm$^{-1}$ (measured on SiO$_2$/Si) within the broadened peak.

The average strain-induced downshift of the A$_1$′ mode is $\delta\omega_{A_1'} = (1.7 \pm 0.2)$ cm$^{-1}$, calculated using $\varepsilon = 1.2\%$, $\omega_{A_1'}^{SiO_2/Si} = (404.0 \pm 0.2)$ cm$^{-1}$, and $\gamma_{A_1'} = 0.18$.[12,13] However, the A$_1$′ mode is visibly split into two components (Fig. 1b), which we define as the lower frequency $A_1'(L)$ mode at $(396.4 \pm 0.3)$ cm$^{-1}$ and the higher frequency $A_1'(H)$ mode at $(403.7 \pm 0.2)$ cm$^{-1}$, corresponding to $\delta\omega_{A_1'(L)} = 7.6$ cm$^{-1}$ and $\delta\omega_{A_1'(H)} = 0.3$ cm$^{-1}$. The most probable origin of the highly downshifted $A_1'(L)$ component is the substrate-induced doping, which affects a portion of the 1L MoS$_2$. The net $A_1'(L)$ shift $\delta\omega_{A_1'(L)}^{corr} = 5.9$ cm$^{-1}$, corrected for the strain by subtracting 1.7 cm$^{-1}$, implies n-type doping of MoS$_2$ and electron concentration estimate of $n_e \sim 2.6 \times 10^{13}$ cm$^{-2}$ for $A_1'(L)$,[14,17] Conversely, the strain-corrected net $A_1'(H)$ shift, $\delta\omega_{A_1'(H)}^{corr} = -1.4$ cm$^{-1}$, points to an electron withdrawal. Since the SiO$_2$/Si reference is known to induce n-doping in MoS$_2$,[18,19] it suggests that $A_1'(H)$ corresponds to regions of undoped MoS$_2$ without a direct contact to Au.

The broadening and splitting of the $E'$ and $A_1'$ Raman modes suggest that the MoS$_2$–Au interaction is heterogeneous, which leads to a multitude of strain and doping states of the MoS$_2$ and results in the convoluted multi-component spectral response observed in Fig. 1b–c. In the extreme case of the top spectrum in Fig. 1b (100 nm Au peeled), the dominating low frequency component $E'(L)$ is accompanied by a high frequency shoulder $E'(H)$, the presence of which is correlated with the $A_1'(H)$ intensity, as discussed below.

The evolution of the Raman spectra with the number of MoS$_2$ layers shown in Fig. 1c for the Au (15 nm e-beam) and SiO$_2$/Si substrates brings further clarity. It transpires that only the bottom-most MoS$_2$ layer interacts strongly with the adjacent Au substrate, while the top layers in 2L, 3L, and bulk MoS$_2$, without a direct contact to Au, are only partially strained and virtually undoped (see also Fig. 4a). We arrive at this conclusion since $E'(L)$ and $A_1'(L)$ are replicated in the thicker layers also, with their frequencies and absolute intensities maintained. These components are readily resolved in 2L, less so in 3L, and negligible in bulk, due to the increased intensities of $E'(H)$ and $A_1'(H)$ (Fig. 1c). Note, that the notation of the Raman modes in MoS$_2$ is layer-dependent due to symmetry considerations. Thus, E′ and $A_1'$ versus $E_{2g}^1$ and $A_{1g}$ irreducible representations are used for an odd number of layers versus even number of layers and bulk.[20]

The biaxial strain and doping observed above are among the highest observed to date.[10,13-15,17,21-23] Previous Raman studies of the $MoS_2$/Au heterostructures either deal with a weak $MoS_2$–Au interaction, are unable resolve the individual spectral components, or offer limited discussion of their origin. Upshifts of both the E′/$E_{2g}^1$ and $A_1'$/$A_{1g}$ modes with the $MoS_2$ thickness were observed using a low-resolution spectrometer,[3,11] consistent with Fig. 1c if heavily averaged spectra are considered. A downshift of E′ to 381 cm$^{-1}$ assigned to strain and an undiscussed low-frequency shoulder near $A_1'$ at 399 cm$^{-1}$ were reported recently.[10] A broadening and downshift of E′ to 379 cm$^{-1}$ due to strain and upshift of the $A_1'$ mode explained by stiffening of the out-of-plane vibration, was observed for 1L $MoS_2$ with the Au deposited on top.[16] This corresponds with the behavior of $A_1'$(H) in our study, but contradicts the appearance of $A_1'$(L).

Both n- and p-type doping of $MoS_2$ interacting with Au have been reported,[10,24,25] probably due to the broad range of work functions ($\Phi$), dependent on preparation, thickness, and surface conditions ($\Phi_{Au}$ ~4.8—5.4 eV,[26,27] $\Phi_{MoS_2}$ ~4.0—5.4 eV[28,29]). This means that the difference $\Delta\Phi = \Phi_{MoS_2} - \Phi_{Au}$ can take both signs. From the ultraviolet photoelectron spectroscopy (UPS) in Supporting Fig. S1 and Kelvin probe force microscopy (KPFM) in Supporting Fig. S2, we estimate $\Delta\Phi$ to be ~0.3–0.2 eV, confirming the net n-type doping of $MoS_2$ induced by the Au.

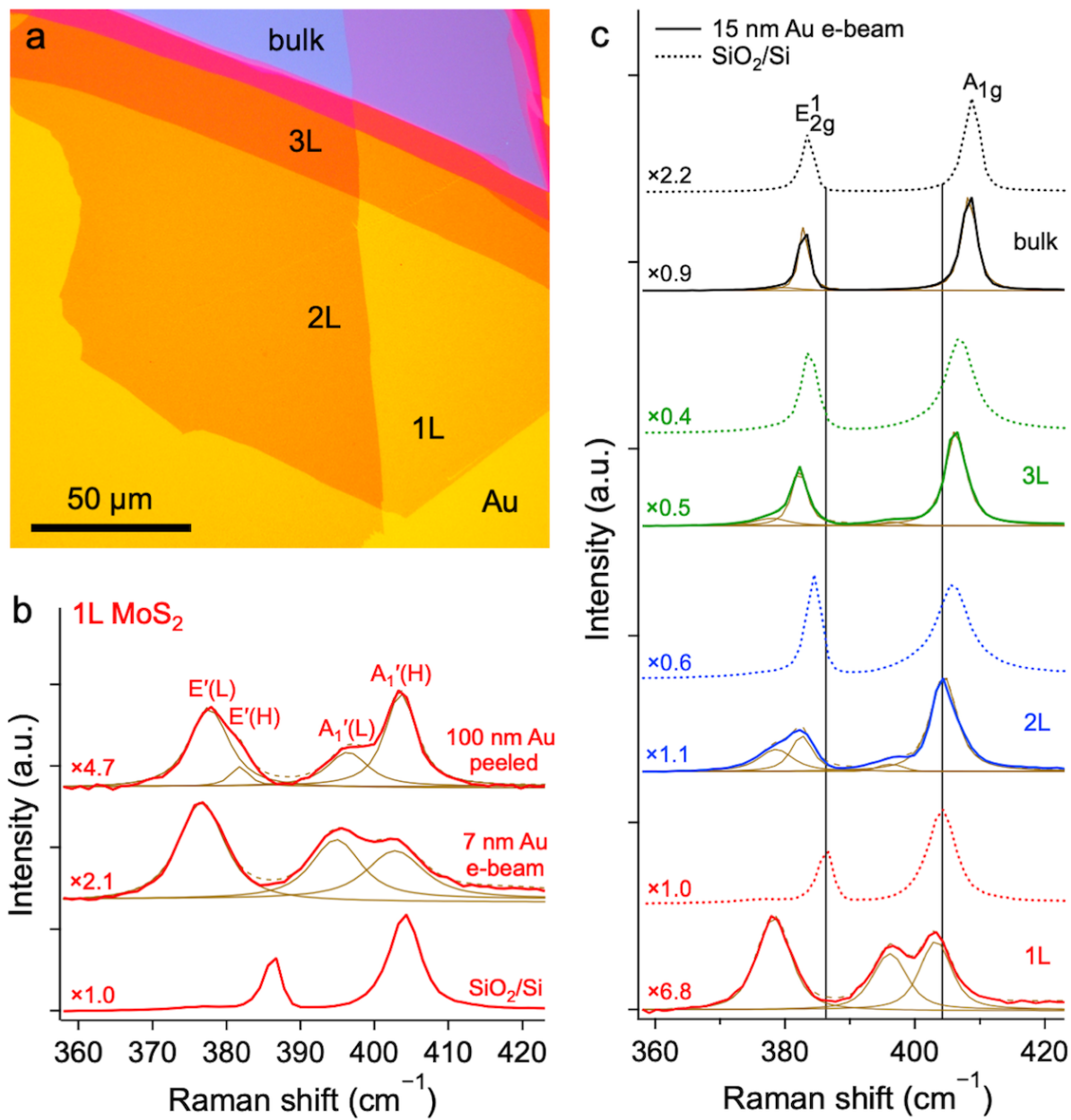

**Figure 1. Far-field micro-Raman spectra of MoS$_2$ on Au. a,** Optical image of MoS$_2$ exfoliated on Au (15 nm e-beam). **b,** Raman spectra of monolayer MoS$_2$ on different substrates: SiO$_2$/Si, 7 nm e-beam Au, and 100 nm thermal Au peeled from Si. **c,** Raman spectra of 1L, 2L, 3L, and bulk MoS$_2$ on 15 nm e-beam Au (solid) and SiO$_2$/Si (dotted). Spectra were collected using a 532 nm excitation and normalized to their highest peaks with the corresponding multiplicators shown on the left. Curve fittings using the Voigt function are shown in brown.

Fig. 2 shows the high-resolution micro-XPS data for the Mo 3d and S 2p core levels, obtained from 1L, 2L, 3L, and bulk $MoS_2$. Photoemission electron microscopy (PEEM) images of the sampled areas are shown in the insets. The Mo 3d and S 2p peaks in 1L are asymmetric and their fitting with two Voigt doublets yields a good match with the spectra, revealing a chemical shift of ~0.4 eV between the higher (H) and lower (L) binding energy components, which appear to have the same origin as $A_1'$(L) and $A_1'$(H) in the 1L Raman spectra (Fig. 1b–c), respectively. The upshift (downshift) of H (L) in 1L from the dominant L component in 2L and 3L (vertical lines) reflects the Fermi level upshift (downshift) in 1L $MoS_2$ due to electron injection (withdrawal). This provides further evidence of the suspected heterogeneity of the $MoS_2$–Au interaction, with n-doped $MoS_2$ in contact with Au (H) and undoped $MoS_2$ detached from Au (L). The H components are also partially replicated in the thicker layers, in analogy to $A_1'$(L) in the Raman spectra of Fig. 1c.

The core level peak energies in bulk $MoS_2$ are less reliable and burdened by larger uncertainties stemming from the weak Au $4f_{7/2}$ signal used as an internal calibration reference, effects of finite probing depth and charging, and presence of step-edges.[30] Importantly, despite the observed shifts in the XPS binding energies and Raman frequencies, the spectral responses are fully consistent with the thermodynamically stable semiconducting 1H phase of 1L $MoS_2$,[3,10] rather than the unstable metallic 1T' phase observed elsewhere.[31,32] This important conclusion demonstrates that the lattice symmetry of 1L $MoS_2$ is preserved and that the Au-induced metallicity can be fully reversed after a transfer onto an insulating substrate.[2,5]

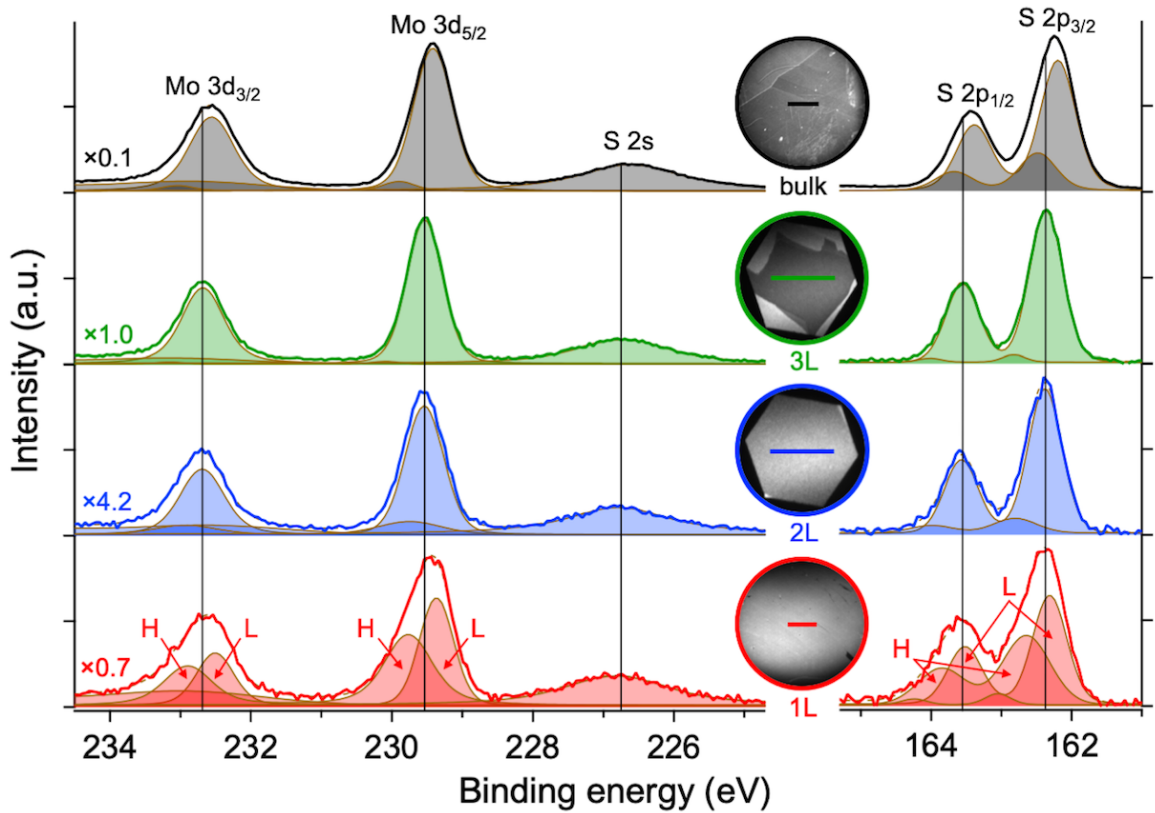

**Figure 2. Micro-XPS of MoS₂ on Au.** Mo 3d (left) and S 2p (right) core level spectra for 1L, 2L, 3L, and bulk MoS$_2$ on Au (15 nm e-beam). Normalization and fitting, qualitatively similar to that of in Fig. 1, were applied. The insets show the PEEM images of the measured areas with 30 μm scale bars.

*Nanoscale Heterogeneity of MoS$_2$ Revealed by TERS*

Near-field TERS with 10 nm spatial resolution allowed us to isolate the A$_1'$(L) and A$_1'$(H) components in the Raman spectra of 1L MoS$_2$ on Au. The optical image in Fig. 3a shows the region used for the TERS mapping in Fig. 3b. The spectra from two individual adjacent (10×10) nm$^2$ pixels shown in Fig. 3c differ greatly and can be matched with the response of 1L MoS$_2$ strongly (S in red) and weakly (W in blue) interacting with the Au. The S spectrum features only A$_1'$(L), while the W spectrum is dominated by A$_1'$(H) with a small A$_1'$(L) shoulder. The blue patches in a larger red region in the right-hand portion of Fig. 3b therefore indicate the presence of weakly interacting nanoscale inclusions in a strongly interacting 1L MoS$_2$ sheet, which appears homogeneous in an optical microscope. TERS signals summed over the pure weakly interacting bilayer (W–2L), pure strongly interacting monolayer (S), and mixed (S+W) regions of the map, shown as solid curves in Fig. 3d, corroborate this conclusion. The mixed region response in particular (magenta in Fig. 3d), is in excellent agreement with the far-field Raman spectrum (dotted curve) recorded in the same region. Unreliability of the absolute intensities in fast TERS mapping make the differentiation between the signals from 2L and weakly interacting 1L challenging. We therefore cannot rule out a disruption of the preferential 1L exfoliation by locally weakened MoS$_2$–Au interaction, potentially leading to nanoscale traces of 2L. In fact, indication of such a phenomenon, with 2L inclusions in a continuous sheet of 1L, was occasionally observed at the microscale (Fig. 3e).

However, a direct proof of the 1L origin of the weakly interacting Raman features is demonstrated by a series of TERS measurements in Fig. 3f, which were acquired using a variable contact tip force on 1L MoS$_2$ transferred onto 50 nm sputtered Au using a polydimethylsiloxane stamp. As no attention was paid to the freshness and cleanliness of the MoS$_2$ and Au surfaces in this case, a layer of contamination was trapped between the two materials, through which we were able to push the TERS tip to alter the distance between the MoS$_2$ and Au, as shown in the inset of Fig. 3f. The top spectrum in Fig. 3f corresponds to the weakly interacting 1L MoS$_2$/Au heterostructure with A$_1'$(H) present but A$_1'$(L) missing. As the tip force increases and the MoS$_2$ is pressed against the Au, their interaction is strengthened and A$_1'$(L) begins to appear at the expense of the A$_1'$(H) intensity.

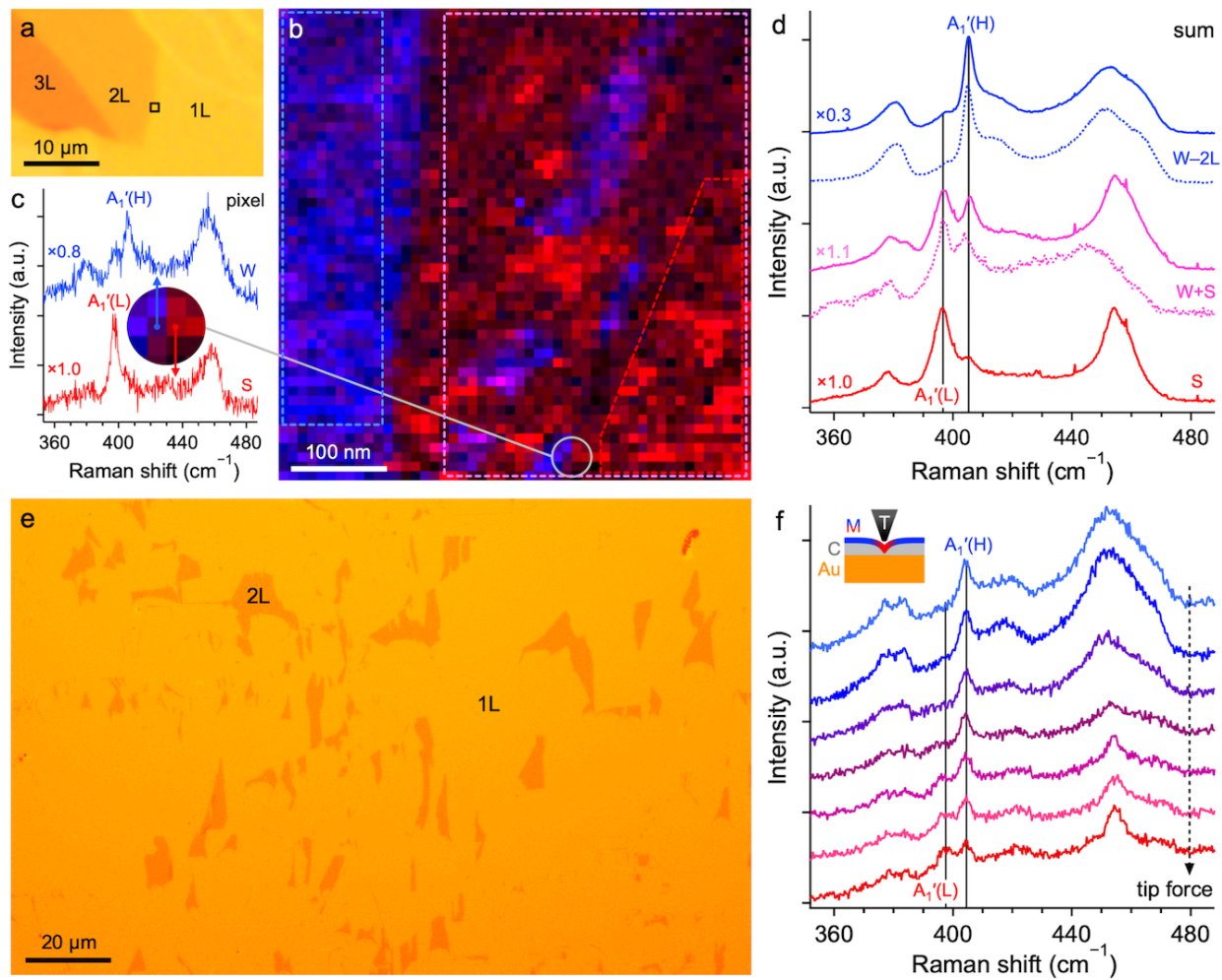

**Figure 3. High-resolution TERS of MoS$_2$ on Au. a,** Optical image of MoS$_2$ on Au (50 nm e-beam). **b,** TERS map (633 nm excitation) of an interface between 1L and 2L defined in **a** by a black rectangle. Red, blue, and magenta hues correspond to the intensities of A$_1'$(L), A$_1'$(H), and both components, respectively. **c,** Single-pixel (10×10) nm$^2$ TERS of the adjacent strongly (S) and weakly (W) interacting 1L MoS$_2$. **d,** TERS summed over the weakly interacting 2L (W–2L), strongly interacting 1L (S), and mixed regions (S+W), from areas highlighted in **b** by dashed polygons of matching colors. Corresponding far-field Raman spectra (633 nm excitation) are shown as dotted curves. **e,** Optical image of microscale 2L inclusions in 1L MoS$_2$. **f,** TERS acquired with a variable tip force in order to alter the MoS$_2$–Au interaction. The inset illustrates how the tip (T) pushes the MoS$_2$ (M) closer to the Au through a layer of contamination (C).

*The Effects of Surface Morphology on Raman Vibrations of MoS$_2$*

A detailed analysis of the far-field Raman data reveals several interesting correlations. Fig. 4a summarizes the evolution of the E′/E$_{2g}^{1}$ and A$_1$′/A$_{1g}$ Raman frequencies with the number of MoS$_2$ layers, for all the Au (color) and SiO$_2$/Si (gray) substrates. The A$_1$′(H) component on Au upshifts with increasing number of layers the same way A$_1$′/A$_{1g}$ does on SiO$_2$/Si.[33,34] In contrast, the A$_1$′(L) component maintains its frequency for 1L–3L, which evidences its origin in the strongly interacting regions of the bottom-most MoS$_2$ layer.

Fig. 4b shows that the ratio between the A$_1$′(L) and A$_1$′(H) intensities of 1L MoS$_2$ [A$_1$′(L)/A$_1$′(H)], proportional to the strength of the MoS$_2$–Au interaction, is strongly correlated with the Au roughness determined by the atomic force microscopy (AFM). This supports the intuitive expectation that the increased conformity of MoS$_2$ to smoother Au surfaces increases strength of their interaction, as schematically depicted in the Fig. 4b insets. In Fig. 4c, we show that A$_1$′(L)/A$_1$′(H) decreases exponentially with the time of Au exposure to air prior to the MoS$_2$ exfoliation. This further evidences the weakening of the MoS$_2$–Au interaction due to airborne contamination (see Fig. 4c insets), in agreement with the observed complete suppression of the initially near-unity 1L yield after 15–20 min of Au exposure to air.[3] For freshly-made Au (day 0), A$_1$′(L)/A$_1$′(H) does not depend on the time elapsed between the Raman measurement and MoS$_2$ exfoliation, in contrast to aged Au (day 7 and 28).

AFM images of the 1L MoS$_2$/Au heterostructure in Fig. 4d–f show signs of MoS$_2$ being suspended between the nanocrystalline features on the Au surface, with a good (poor) contact at the protrusions (depressions). As the Au roughness increases, larger areas of MoS$_2$ decouple from the substrate, which is reflected by the increased intensity of the weakly interacting A$_1$′(H) component (insets of Fig. 4b). Larger proportion of suspended MoS$_2$ with a weak MoS$_2$–Au interaction also leads to a relative increase in the E′(H) intensity, as seen for the roughest Au substrate in Fig. 1b (top spectrum). The proportional increase of the strongly interacting MoS$_2$–Au regions through an increased contact area is further evidenced by an Au/MoS$_2$/Au "sandwich" prepared by covering 1L MoS$_2$ on 15 nm Au with another layer of 5 nm Au. This leads to a 40% increase in $A_1′(L)/A_1′(H)$ indicated by the gray-filled marker in Fig. 4b.

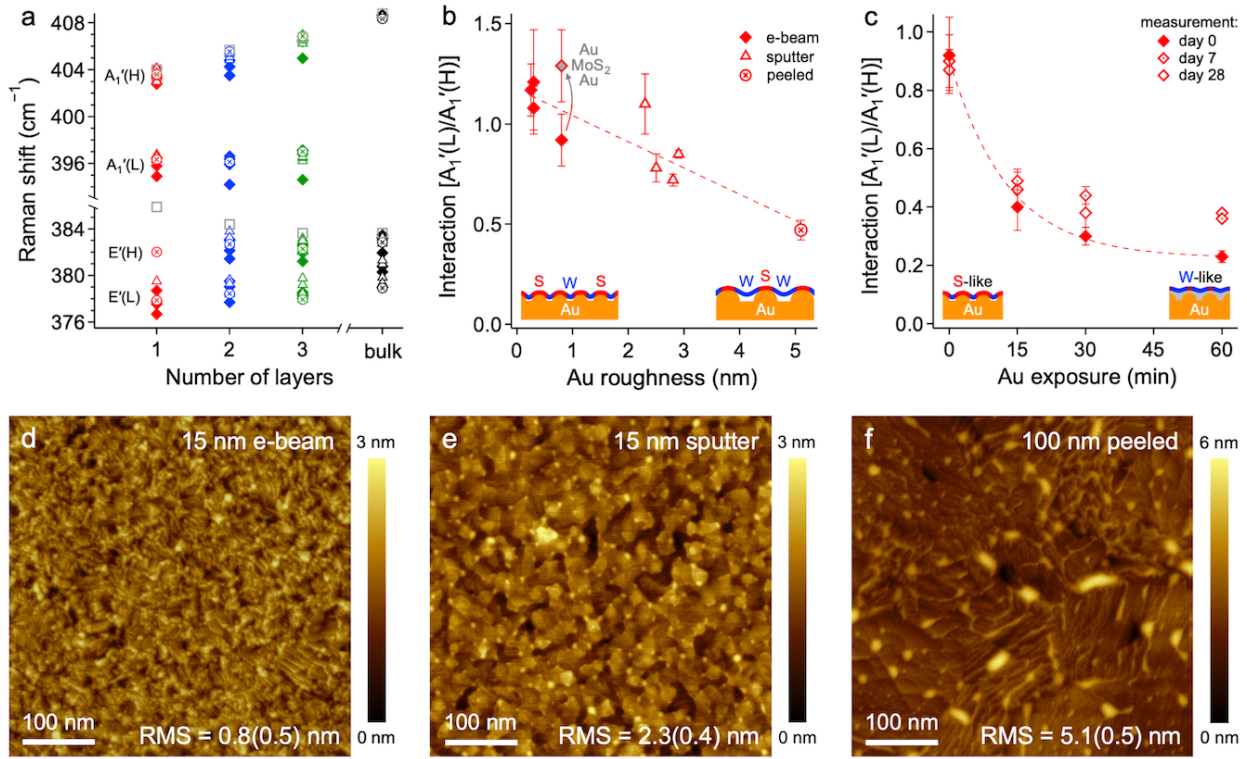

**Figure 4. The effects of surface morphology on Raman vibrations of MoS$_2$. a,** Raman frequencies as a function of the number of MoS$_2$ layers for all measured samples. **b,** A$_1'$(L)/A$_1'$(H) as a function of the Au roughness, determined by AFM. In comparison, the average roughness of 1L MoS$_2$ was (0.4 ± 0.1) nm. The gray-filled marker corresponds to the 5 nm Au/1L MoS$_2$/15 nm Au "sandwich". **c,** A$_1'$(L)/A$_1'$(H) as a function of the Au exposure to air prior to the MoS$_2$ exfoliation (0, 15, 30, 60 min), measured at different times after sample preparation (day 0, 7, 28). Diamond, triangle, circle, and square markers in **a–c,** obtained using the Voigt fitting of the spectra, denote the e-beam Au, sputtered Au, peeled Au, and SiO$_2$/Si substrates, respectively. **d–f,** AFM images of 1L MoS$_2$ on 15 nm e-beam, 15 nm sputter, and 100 nm peeled Au, respectively, noting the root mean square roughness (RMS) of the Au (MoS$_2$) surface.

*Origin of the A₁′ mode splitting*

The experimental evidence presented in this study unequivocally links the appearance of the downshifted $A_1'(L)$ Raman mode to that portion of 1L MoS$_2$, which strongly interacts with its Au substrate. The n-doping of 1L MoS$_2$ in contact with Au, proven by XPS, UPS, and KPFM, corroborates that the increased electron concentration is responsible for the $A_1'(L)$ downshift, in line with electrochemically-gated MoS$_2$ experiments.[14,15,17] Nevertheless, one could envisage alternative explanations. The strong binding in MoS$_2$–Au heterostructure with a clean interface could cause softening of the Mo–S bonds, instead of the stiffening seen for contaminated MoS$_2$–Au interface or bulk MoS$_2$.[16,33] Alternatively, the strong interaction could lead to activation of phonons otherwise silent in 1L MoS$_2$, such as those present in multilayer systems.[34,35] However, this option can be ruled out for $A_1'(L)$, since the activation of another mode would not lead to a disappearance of the original $A_1'$, in contrast to our TERS results. We also examined a 1L WS$_2$/Au heterostructure prepared by the same technique and observed the same apparent splitting of $A_1'$ into $A_1'(L)$ and $A_1'(H)$ in the far-field Raman spectra, with the $A_1'(L)$ downshifted by ~7 cm$^{-1}$ (Supporting Fig. S3). This is a nearly identical shift to that for 1L MoS$_2$, which clearly points to an electron density increase rather than activation of a new mode.

**CONCLUSIONS**

We studied Raman spectroscopy and XPS of $MoS_2$ on gold and identified the specific vibrational and binding energy fingerprints of the strong $MoS_2$–Au interaction. Far-field micro-Raman reveals significant downshift and broadening of the in-plane E′ mode of 1L $MoS_2$ on Au, compared to $MoS_2$ on $SiO_2$/Si, which corresponds to heterogeneous tensile biaxial strains of up to 1.9%. Splitting of the out-of-plane $A_1$′ mode of 1L $MoS_2$ into two separate components implies that a portion of $MoS_2$ in close contact with Au experiences n-type charge transfer doping with electron concentrations up to $2.6 \times 10^{13}$ $cm^{-2}$, while another portion of $MoS_2$ is suspended and remains undoped. This is supported by splitting in the XPS of the Mo 3d and S 2p core levels. The evolution of the micro-Raman spectra and XPS with the $MoS_2$ thickness confirms that the strong $MoS_2$–Au interaction is confined to the bottom-most $MoS_2$ layer. High-resolution TERS mapping confirms the suspected nanoscale heterogeneity of the $MoS_2$–Au interaction caused by the spatial non-conformity between the two materials. Finally, the micro-Raman data show that the $MoS_2$–Au interaction can be effectively tuned by the surface morphology and cleanliness of the underlying Au substrate, which could be exploited for strain and charge doping engineering of $MoS_2$ and utilization of the induced metallicity of gold-exfoliated TMDCs in catalysis.

**METHODS**

**Sample Preparation**

Gold films on 90 nm SiO$_2$/Si wafers (IDB Technologies Ltd) were prepared by three different methods: magnetron sputtering (CMS-A, Kurt J Lesker Company Ltd), e-beam evaporation (SC4500, CVC Products Inc), and thermal evaporation (DV502-A, Denton Vacuum Inc) followed by peeling from a sacrificial Si substrate.[8] Au thicknesses ranging from 3 to 100 nm were prepared, and an adhesion layer of 1 nm or 3 nm Ti was employed for the sputtered and e-beam samples, respectively. MoS$_2$ was exfoliated onto the Au surface from bulk molybdenite crystals (Manchester Nanomaterials Ltd), using a low-stain tape.

**Characterization**

The exfoliated MoS$_2$ was inspected, and its thickness determined, using a Nikon L200N Eclipse optical microscope. MFP-3D AFM (Asylum Research) in tapping mode was used to determine the surface roughness. An Icon Dimension AFM (Bruker Corp.) in PeakForce tapping mode using Scanasyst-Air probes was employed for the high-resolution characterization of the surface. Far-field Raman spectra were collected using an inVia Reflex confocal spectrometer (Renishaw plc) with a 532 nm laser and 2400 l/mm grating and LabRAM HR (Horiba Ltd) with a 633 nm laser and 1800 l/mm grating, focused to ~1 μm$^2$ spot size by a 100× objective. Near-field TERS was measured using a LabRAM Nano system comprised of HR Evolution spectrometer and OmegaScope-R SPM (HORIBA Scientific) with a 633 nm laser, 1800 l/mm grating, and Ag-coated Si tips (App Nano), using 1 s (3 s) integration time for mapping (tip force) measurements, respectively, and <300 μW laser power for each pixel. XPS, UPS and PEEM were measured in NanoESCA microscope (Omicron). The XPS was collected using a monochromated Al Kα source (hν = 1486.7 eV), and the UPS was carried out using He I discharge lamp (hν = 21.2 eV). The XPS calibration was done using the Au 4f$_{7/2}$ core level at 84 eV along with the Fermi level edge.

## ASSOCIATED CONTENT

**Supporting Information:**

The Supporting Information is available free of charge on the ACS Publication website at DOI:

Work function estimation from UPS; work function estimation from KPFM; Raman spectra of 1L $WS_2$ on Au.

## AUTHOR INFORMATION
**Corresponding authors**
*matej.velicky@manchester.ac.uk
*otakar.frank@jh-inst.cas.cz

## ACKNOWLEDGEMENTS

This project has received funding from the European Union's Horizon 2020 research and innovation program under the Marie Skłodowska-Curie grant agreement No. 746685 and Czech Science Foundation project GACR 17-18702S. This work was performed in part at the Cornell NanoScale Science & Technology Facility, a member of the National Nanotechnology Coordinated Infrastructure, which is supported by the NSF (Grant NNCI-1542081), and made use of the Cornell Center for Materials Research Shared Facilities, which are supported through the NSF MRSEC program (DMR-1719875). This work was also supported by the Ministry of Education, Youth and Sports of the Czech Republic and The European Union - European Structural and Investments Funds in the frame of Operational Programme Research Development and Education project Pro-NanoEnviCz (Reg. No. CZ.02.1.01/0.0/0.0/16_013/0001821). Photoemission spectroscopy experiments were supported by MEYS of the Czech Republic through project CZ.02.1.01/0.0/0.0/16_013/0001406.

**Table of Contents (TOC):**

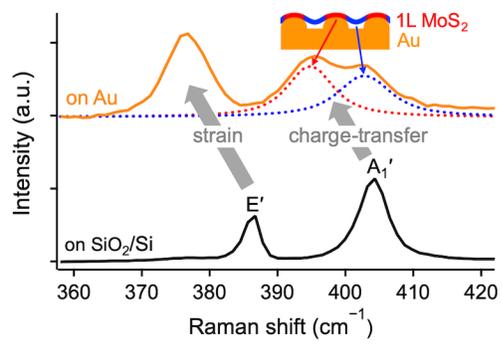

# Fingerprints of the Strong Interaction between Monolayer MoS$_2$ and Gold

Matěj Velický,*[1,2,3] Alvaro Rodriguez,[4] Milan Bouša,[4] Andrey V. Krayev,[5] Martin Vondráček,[6] Jan Honolka,[6] Mahdi Ahmadi,[2] Gavin E. Donnelly,[3] Fumin Huang,[3] Héctor D. Abruña,[2] Kostya S. Novoselov,[1,7,8] and Otakar Frank*[4]

[1] Department of Physics and Astronomy, University of Manchester, Oxford Road, Manchester, M13 9PL, United Kingdom

[2] Department of Chemistry and Chemical Biology, Cornell University, Ithaca, New York, 14853, United States

[3] School of Mathematics and Physics, Queen's University Belfast, University Road, Belfast, BT7 1NN, UK

[4] J. Heyrovský Institute of Physical Chemistry, Czech Academy of Sciences, Dolejškova 2155/3, 182 23 Prague, Czech Republic

[5] HORIBA Scientific, Novato, CA, 94949, United States

[6] Institute of Physics, Czech Academy of Sciences, Na Slovance 1999/2, 182 21 Prague 8, Czech Republic

[7] Centre for Advanced 2D Materials, National University of Singapore, 117546, Singapore

[8] Chongqing 2D Materials Institute, Liangjiang New Area, Chongqing, 400714, China

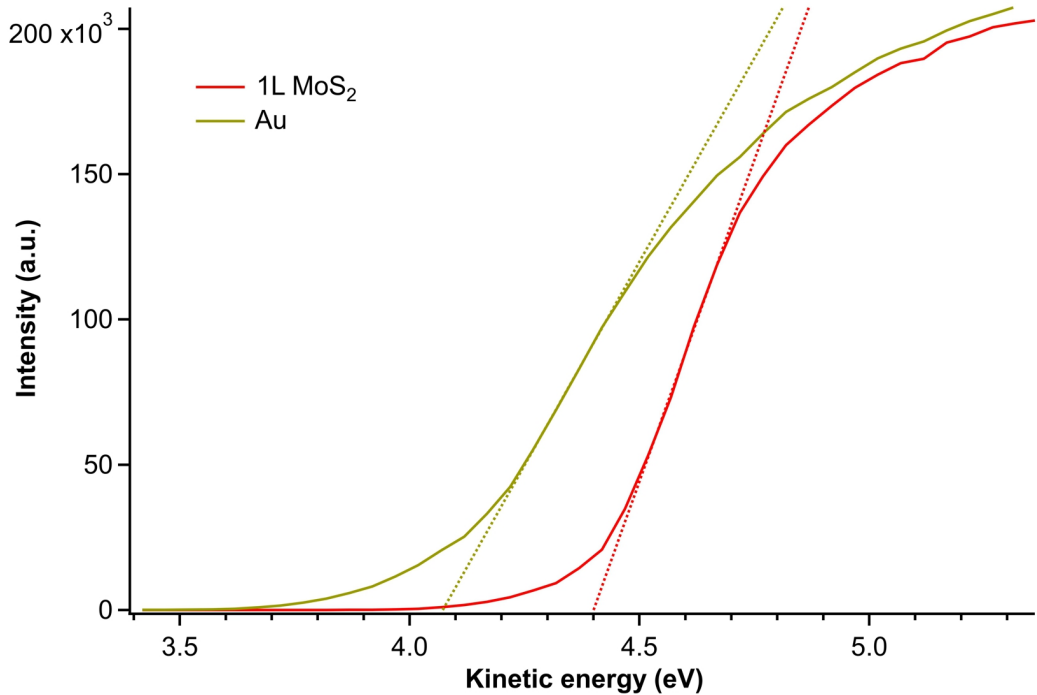

**Supporting Figure S1. Work function estimation using ultraviolet photoelectron spectroscopy.** Low kinetic energy cut-off spectra of 1L MoS$_2$ on Au (solid red; 15 nm e-beam) and of bare Au substrate (solid green) measured next to the MoS$_2$, with the linear fits (dashed) through the inflection points of the curves. The work function ($\Phi$) is determined from the value of the fitted line at zero intensity. $\Phi_{MoS_2}$ was found to be ~0.3 eV larger in comparison to $\Phi_{Au}$. Although the absolute $\Phi$ values are burdened with an uncertainty due to the spectrometer response, their difference yields a reliable work function difference estimate.

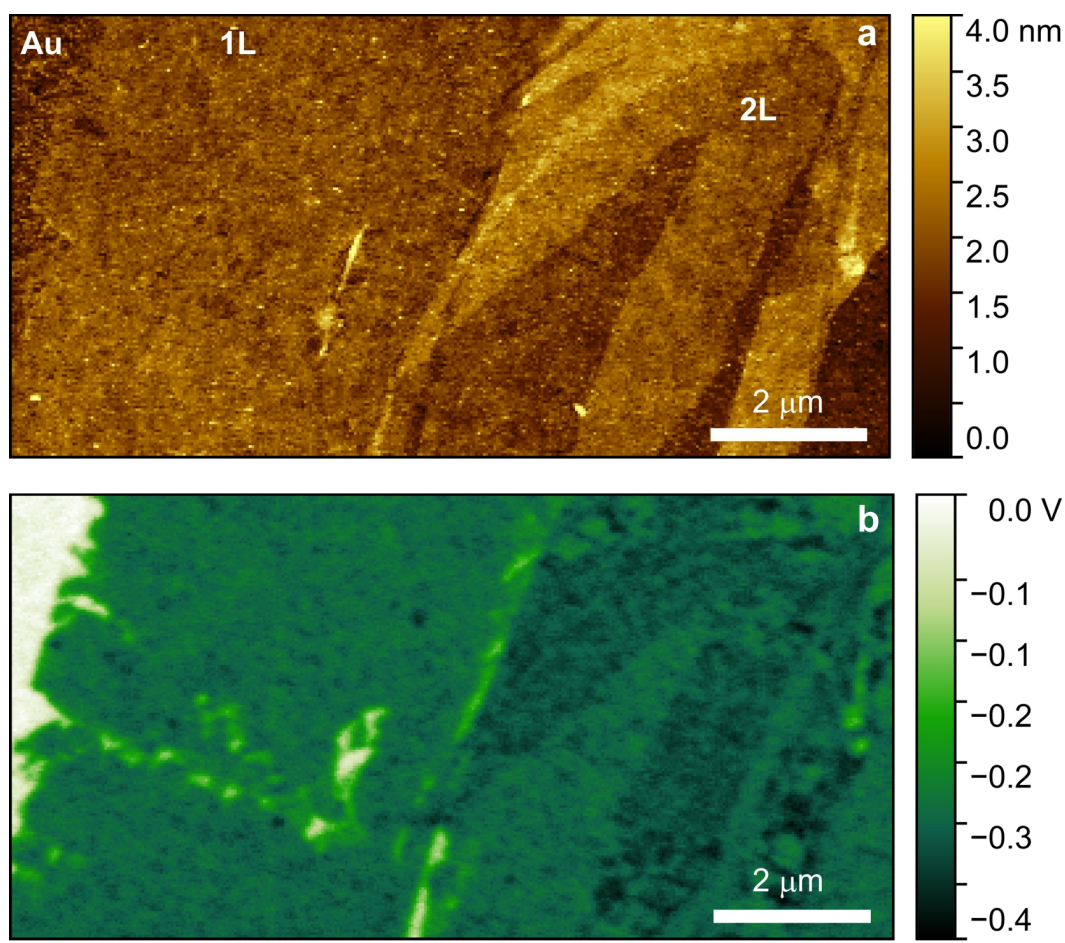

**Supporting Figure S2. Work function estimation using Kelvin probe force microscopy. a,** Topography atomic force microscopy image of an area containing 1L and 2L MoS$_2$ and the bare Au substrate (50 nm e-beam). **b,** Contact potential difference (CPD) map of **a** obtained by Kelvin probe force microscopy. The CPD of 1L MoS$_2$ is lower, and therefore its $\Phi$ is larger, than that of the Au substrate. The corresponding difference in the CPD of the two materials, averaged over the whole measured area, is 0.22 V. The trend of increasing work function (decreasing CPD) with the number of layers observed here for 1L/2L MoS$_2$ is consistent with other literature observations.[1,2]

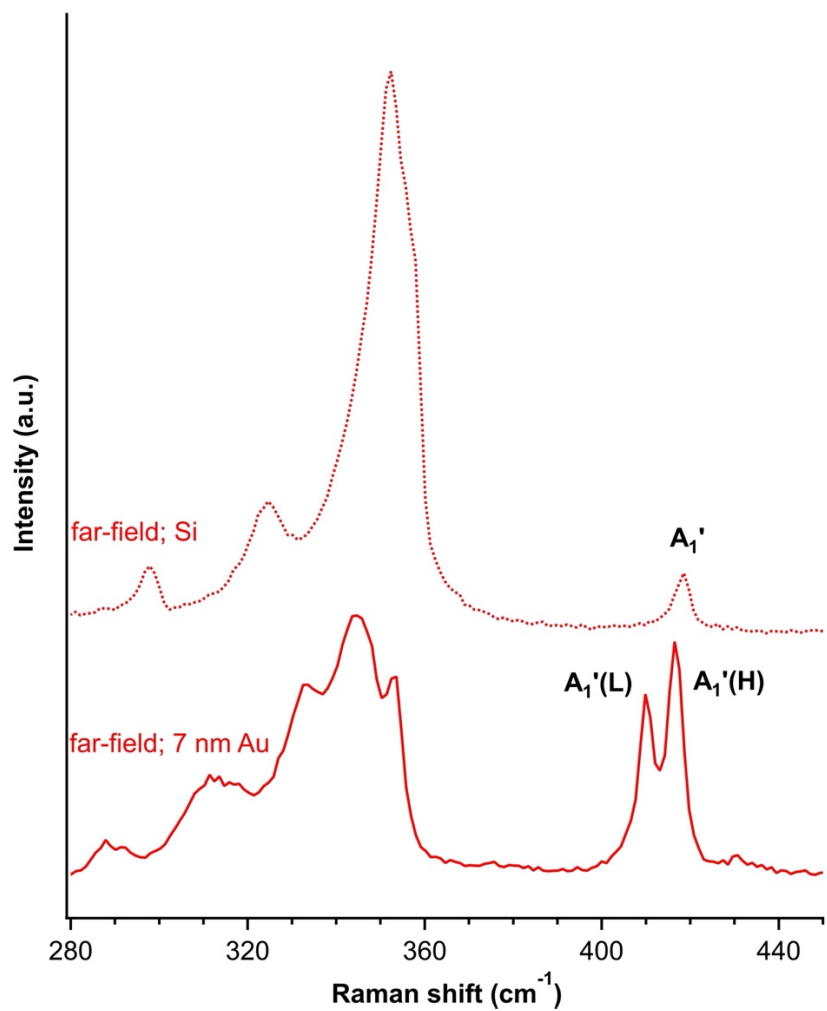

**Supporting Figure S3. Raman spectra of 1L WS$_2$.** Far-field micro-Raman spectra of 1L WS$_2$ on 7 nm sputtered Au (bottom curve) and on Si/SiO$_2$ (top curve), collected using 532 nm excitation.

**Supporting References**